# Zero-shot Multi-Contrast Brain MRI Registration by Intensity Randomizing T1-weighted MRI (LUMIR25)

Hengjie Liu[1,2], Yimeng Dou[3], Di Xu[1], Xinyi Fu[1], Dan Ruan[2] and Ke Sheng[1]

[1] University of California, San Francisco, San Francisco CA, USA
[2] University of California, Los Angeles, Los Angeles CA, USA
[3] University of Wisconsin, Madison, Madison WI, USA

**Abstract.** In this paper, we present our submission to the LUMIR25 task of Learn2Reg 2025, which ranked 1st overall on the test set. Extended from LUMIR24, this year's task focuses on zero-shot registration under domain shifts (e.g., high-field MRI, pathological brains, and various MRI contrasts), while the training data comprises only in-domain T1-weighted brain MRI. We start with a meticulous analysis of LUMIR24 winners to identify the main contributors to strong monomodal registration performance. We highlight the importance of registration-specific inductive biases, including multi-resolution pyramids, inverse and group consistency, topological preservation or diffeomorphism, and correlation-based correspondence establishment. To further generalize to diverse contrasts, we employ three simple but effective strategies: (i) a multimodal loss based on the modality-independent neighborhood descriptor (MIND), (ii) intensity randomization for unseen contrast augmentation, and (iii) lightweight instance-specific optimization (ISO) on feature encoders at inference time. On the validation set, the proposed approach substantially improves T1–T2 registration accuracy, demonstrating robust cross-contrast generalization without relying on explicit image synthesis. These results suggest a practical step toward a registration foundation model that can leverage a single training domain yet remain robust across domain shifts.

## 1 Introduction

The LUMIR24 challenge introduced a large-scale dataset for monomodal T1-weighted brain MRI registration [1,2] and yielded fruitful results demonstrating the strength of deep learning in deformable image registration [3]. The winning method, SITReg [4], succeeded by emphasizing registration-specific inductive biases, including multi-resolution pyramids, by-construction inverse consistency (IC), group consistency (GC) and topological preservation, without resorting to complicated network architectures or advanced computation blocks such as Transformers or Mamba. The best-performing baseline, vector field attention (VFA) [5], was also notable for using a deterministic module to extract displacement directly from correlation features. Their success resonates with the recent studies that argue registration-specific designs matter more than choices of computation blocks [6–9]. Building on these insights, we closely examined the LUMIR24 leaders to identify the key "recipe" for strong monomodal registration.

As LUMIR25 shifts toward a foundation model capable of zero-shot registration while being trained only on T1-weighted images, we adopt three simple yet effective strategies to handle multimodal settings: (i) a multimodal loss based on the modality-independent neighborhood descriptor (MIND) [10]; (ii) intensity augmentation using smooth randomized mappings, and (iii) lightweight instance-specific optimization (ISO) applied only to the feature encoders with deformation prediction modules frozen.

## 2   Methods

### 2.1   Key components for monomodal registration (LUMIR24)

We start by investigating key registration-specific design choices that are critical to registration performance inspired by [4–7, 11]. These include the multiresolution pyramid, correlation calculation, and inverse consistency. We design a unified framework to test the contribution of each component, as detailed in Figure 1 and Table 1. Additional implementation details can be found in [8] and in our code repository: https://github.com/HengjieLiu/Unsupervised-DL-DIR-Revisited.

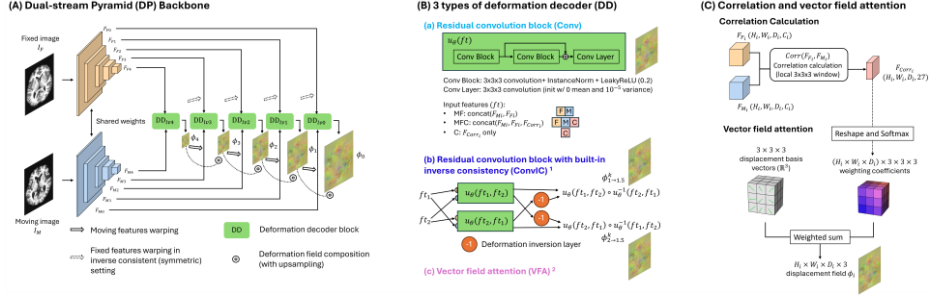

**Fig. 1.** (A) The standardized dual-stream pyramid (DP) backbone for registration. (B) Three types of deformation decoders. (C) Illustration of correlation calculation and vector field attention.

The registration results on LUMIR24 validation set are shown in Table 2. The multi-resolution pyramid is key to achieving significantly better accuracy compared with VoxelMorph [12] and TransMorph [13] baselines. The inverse-consistent (IC) variants further improve regularity, as indicated by a lower non-diffeomorphic volume (NDV) [14]. Further, the correlation-only variants, despite having fewer parameters, surpass their counterparts, highlighting the role of correlation in deformation estimation.

However, in practice, correlation layers are memory intensive. Under our GPU constraint (48 GB VRAM), scaling up a backbone without correlation outperformed a smaller correlation-based model. In addition, the group consistency (GC) loss and the NDV loss—introduced by last year's winner—were highly effective at reducing HD95 and NDV, respectively, as shown in Table 3. Consequently, we still base our final model on the original SITReg without using correlation (similar to (b) DP-ConvIC-MF), and incorporates GC and NDV losses.

Table 1. High level summary of proposed methods and comparing methods.

| Methods | Coarse-to-fine | Inverse consistency | Input features | Params (M) (encoder/decoder) |
|---|---|---|---|---|
| VoxelMorph | ✗ | ✗ | F M | 0.3 |
| TransMorph | ✗ | ✗ | F M | 46.56 |
| VFA | ✓ | ✗ | C | 2.01 |
| SITReg | ✓ | ✓ | F M | 15.08 |
| (a) DP-Conv-MF | ✓ | ✗ | F M | (0.51/2.35) 2.85 |
| (a) DP-Conv-MFC | ✓ | ✗ | F M C | (0.51/2.66) 3.17 |
| (a) DP-Conv-C | ✓ | ✗ | C | (0.51/1.49) 1.99 |
| (b) DP-ConvIC-MF | ✓ | ✓ | F M | (0.51/2.35) 2.85 |
| (b) DP-ConvIC-C | ✓ | ✓ | C | (0.51/1.80) 2.31 |
| (c) DP-VFA | ✓ | ✗ | C | (0.51/0.28) 0.79 |

Table 2. Results on LUMIR24 validation set.

| Methods | Dice ↑ | HD95 ↓ | TRE ↓ | NDV (%) ↓ |
|---|---|---|---|---|
| VoxelMorph | 0.7186 ± 0.0340 | 3.9821 | 3.1545 | 1.1836 |
| TransMorph | 0.7594 ± 0.0319 | 3.5074 | 2.4225 | 0.3509 |
| (a) DP-Conv-MF | 0.7713 ± 0.0290 | 3.3534 | 2.4676 | 0.4158 |
| (a) DP-Conv-MFC | 0.7730 ± 0.0291 | 3.3566 | 2.4449 | 0.4672 |
| (a) DP-Conv-C | 0.7747 ± 0.0295 | 3.3666 | 2.4135 | 0.3795 |
| (b) DP-ConvIC-MF | 0.7717 ± 0.0288 | 3.3489 | 2.3660 | 0.0310 |
| (b) DP-ConvIC-C | 0.7724 ± 0.0288 | 3.3873 | 2.3357 | 0.0309 |
| (c) DP-VFA | 0.7764 ± 0.0284 | 3.2157 | 2.4420 | 0.0540 |
| VFA | 0.7726 ± 0.0286 | 3.2127 | 2.4949 | 0.0788 |
| SITReg | 0.7742 ± 0.0291 | 3.3039 | 2.3112 | 0.0231 |
| SITReg (GC/NDV) | **0.7805 ± 0.0287** | **3.1187** | **2.3005** | **0.0025** |

### 2.2 Extension to multimodal registration (LUMIR25)

**MIND loss.** Our first adaptation for multimodal registration is to use a MIND-based similarity loss [10] instead of normalized cross correlation (NCC). The loss function is

$$Loss = \lambda_1 L_{sim} + \lambda_2 L_{smooth} + \lambda_3 L_{GC} + \lambda_4 L_{NDV}, \quad (1)$$

where $\lambda_1 = 1$ for NCC and $\lambda_1 = 10$ for MIND, $\lambda_2 = 1$ with a diffusion regularizer, and $\lambda_3 = 40$ and $\lambda_4 = 1e-5$ as in SITReg. Although all LUMIR24 participants used NCC loss [3], we find that MIND performs competitively with NCC on LUMIR24 (Table 3). Notably, it improves TRE, which is expected since MIND is sensitive to edge and corner structures. Adding GC and NDV losses further improves accuracy (Dice and HD95) for MIND as well.

Table 3. Comparison of NCC loss vs MIND loss on LUMIR24 validation set.

| Methods | Dice ↑ | HD95 ↓ | TRE ↓ | NDV (%) ↓ |
| --- | --- | --- | --- | --- |
| SITReg-NCC | 0.7735 ± 0.0285 | 3.3183 | 2.3266 | 0.0267 |
| SITReg-NCC (GC/NDV) | **0.7798 ± 0.0289** | 3.1296 | 2.3102 | **0.0024** |
| SITReg-MIND | 0.7705 ± 0.0276 | 3.3781 | **2.2702** | 0.1045 |
| SITReg-MIND (GC/NDV) | 0.7761 ± 0.0290 | **3.0685** | 2.3139 | 0.0067 |

**Intensity augmentation.** Intensity augmentation has been widely used in domain generalization for medical image segmentation [15]. To mimic inter-sequence appearance shifts while preserving anatomical structures, we apply a smooth, randomized pointwise intensity remapping to each T1-weighted training volume. Specifically, we define a smooth intensity remapping function $g(x)$ using a shape-preserving piecewise-cubic Hermite interpolant (PCHIP), which yields a $C^1$ mapping. We parameterize $g$ by $n_{knots}$ control points $\{(x_i, y_i)\}_{i=1}^{n_{knots}}$ over the intensity range $[0, 255]$, where the knot locations $x_i$ are uniformly spaced. We fix the endpoints to satisfy $g(0) = 0$ and $g(255) = 255$, and randomly sample the remaining $n_{knots} - 2$ interior values $y_i$. We then discretize $g$ at integer intensities to obtain a 256-entry lookup table $T: \{0, \ldots, 255\} \to \{0, \ldots, 255\}$, and apply it voxelwise via $v' = T(v)$. To avoid degenerate contrast collapse, we reject candidates whose post-mapping histograms show saturated bins with a preset threshold. We empirically set $n_{knots} = 6$, but further optimization could yield additional gains. Examples of augmented images are shown in Figure 2. The augmented images can resemble other contrasts such as T2-weighted images. In total, we precomputed 2,000 mappings and applied them randomly during training.

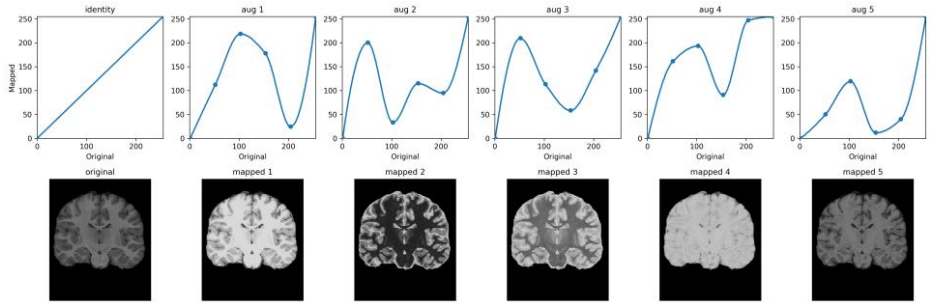

**Fig. 2.** Intensity randomization examples for multimodal emulation. Aug #4 will be rejected.

**ISO strategy.** In LUMIR24, SITReg outperformed other methods on in-domain T1–T1 registration without relying on instance-specific optimization (ISO). However, ISO has been reported to enhance robustness for zero-shot tasks [3, 16]. While ISO can further decrease the pairwise registration loss at inference time, it may overfit to intensity matching (i.e., the similarity loss). Empirically, we observed better Dice and HD95 with strong GC regularization, whereas pairwise ISO can break away from this regularization and slightly worsen Dice and HD95. Therefore, we did not use ISO for T1–T1 tasks and experimented with ISO for other modalities. To mitigate overfitting, we apply ISO only to the feature encoder (ISO-fe) while keeping the deformation decoder frozen. This is reasonable because the decoder has already been exposed to diverse feature styles through augmentation, whereas adapting the encoder can better accommodate unseen intensity profiles while limiting the risk of overfitting. Moreover, ISO-fe updates only 22% of the total parameters compared with ISO-full. We ran both ISO-fe and ISO-full for 20 steps using the loss function in Eq. 1, while omitting the GC term.

### 2.3 Submitted method

Our final submission integrates two models to maximize performance under the challenge evaluation metrics. For T1–T1 registration, we use SITReg-NCC (GC/NDV) without ISO, as it performs slightly better than SITReg-MIND (GC/NDV). For all other imaging contrast pairs, we use SITReg-MIND-Aug (GC/NDV), with ISO-fe applied at inference time.

The models were implemented in PyTorch 2.5.1, and all training was conducted on NVIDIA RTX 6000 Ada GPUs with 48 GB of memory. Group consistency training required three GPUs to process three image pairs in parallel, whereas ISO was performed on a single GPU.

## 3 Results

The validation set comprises 36 registration pairs in total:
- 18 in-domain (ID) T1–T1 pairs (9 evaluated with contour-based metrics and 9 with landmark-based metrics),
- 9 out-of-domain (OD) high-field T1–T1 pairs (evaluated with contour-based metrics),
- 9 multimodal (MM) T1–T2 pairs (evaluated with contour-based metrics).

### 3.1 Ablation on proposed strategies

Table 4 reports the validation results on the three subsets (ID, OD, and MM), presenting an ablation study that highlights the contribution of each component. The key findings are as follows.

**Table 4.** Registration results on the validation set across three subsets, illustrating the effects of the proposed components: MIND similarity, intensity augmentation (Aug), and instance-specific optimization (ISO). All SITReg variants are trained with GC and NDV losses.

(a)  In-domain T1 – T1 (9 contour-based+9 landmark-based=18 pairs)

| Methods | Dice ↑ | HD95 ↓ | TRE ↓ | NDV (%) ↓ |
| --- | --- | --- | --- | --- |
| SynthMorph (Baseline) | 0.7262 ± 0.0133 | 3.5763 ± 0.1963 | 2.6099 ± 0.3627 | 0.0000 ± 0.0000 |
| SITReg-NCC | **0.7816 ± 0.0171** | 3.1091 ± 0.2445 | 2.3103 ± 0.2705 | 0.0023 ± 0.0005 |
| SITReg-MIND | 0.7796 ± 0.0170 | 3.1102 ± 0.2363 | 2.3358 ± 0.3165 | 0.0075 ± 0.0008 |
| SITReg-MIND-Aug | 0.7791 ± 0.0159 | **3.0974 ± 0.2339** | 2.2988 ± 0.2993 | 0.0069 ± 0.0009 |
| SITReg-MIND-Aug + ISO-full | 0.7722 ± 0.0156 | 3.3213 ± 0.2695 | **2.2615 ± 0.3068** | 0.0047 ± 0.0006 |
| SITReg-MIND-Aug + ISO-fe | 0.7730 ± 0.0152 | 3.2721 ± 0.2488 | 2.2948 ± 0.3024 | 0.0059 ± 0.0009 |

(b)  Out-of-domain (high-field) T1–T1 (9 pairs)

| Methods | Dice ↑ | HD95 ↓ | NDV (%) ↓ |
| --- | --- | --- | --- |
| SynthMorph (Baseline) | 0.6888 ± 0.0202 | 3.9061 ± 0.4199 | 0.0000 ± 0.0000 |
| SITReg-NCC | **0.7652 ± 0.0157** | 3.1738 ± 0.3244 | 0.0027 ± 0.0001 |
| SITReg-MIND | 0.7610 ± 0.0148 | **3.1648 ± 0.3097** | 0.0079 ± 0.0008 |
| SITReg-MIND-Aug | 0.7595 ± 0.0145 | 3.2032 ± 0.2845 | 0.0077 ± 0.0006 |
| SITReg-MIND-Aug + ISO-full | 0.7575 ± 0.0119 | 3.3525 ± 0.2937 | 0.0069 ± 0.0008 |
| SITReg-MIND-Aug + ISO-fe | 0.7560 ± 0.0134 | 3.3192 ± 0.3296 | 0.0065 ± 0.0009 |

(c)  Multimodal T1 – T2 (9 pairs)

| Methods | Dice ↑ | HD95 ↓ | NDV (%) ↓ |
| --- | --- | --- | --- |
| SynthMorph (Baseline) | 0.6888 ± 0.0244 | 3.2489 ± 0.3091 | 0.0000 ± 0.0000 |
| SITReg-NCC | 0.3564 ± 0.0192 | 6.0990 ± 0.3403 | 0.0007 ± 0.0003 |
| SITReg-MIND | 0.3691 ± 0.0175 | 5.9121 ± 0.4434 | 0.0097 ± 0.0021 |
| SITReg-MIND-Aug | 0.7165 ± 0.0238 | **2.7409 ± 0.1777** | 0.0044 ± 0.0008 |
| SITReg-MIND-Aug + ISO-full | 0.7236 ± 0.0276 | 2.8723 ± 0.1907 | 0.0049 ± 0.0007 |
| SITReg-MIND-Aug + ISO-fe | **0.7241 ± 0.0284** | 2.8328 ± 0.1977 | 0.0048 ± 0.0007 |

**MIND vs NCC.** NCC remains competitive for monomodal T1–T1 registration. For both in-domain and out-of-domain T1–T1, SITReg-NCC achieves the best Dice, while differences in HD95 and TRE vary across methods and require further validation on larger test sets. Using MIND loss with ISO can improve landmark-related alignment (TRE), indicating better point-wise correspondence; however, this does not translate into higher Dice, suggesting a potential trade-off between landmark fidelity and volumetric overlap.

**Intensity augmentation.** Intensity augmentation is essential for cross-contrast generalization. Without augmentation, both SITReg-NCC and SITReg-MIND perform poorly on multimodal T1–T2 registration. Introducing the proposed intensity randomization markedly improves robustness. SITReg-MIND-Aug shows strong T1–T2 performance and consistently outperforms the SynthMorph [17] baseline across all three validation subsets, highlighting the effectiveness of learning from real T1 anatomy while simulating diverse contrast appearances.

**ISO for T1–T1.** ISO is not beneficial for T1–T1. Applying ISO to monomodal T1–T1 tends to slightly degrade Dice and HD95 in both in-domain and out-of-domain settings, suggesting that pairwise ISO may overfit to the similarity objective and partially undermine the benefits induced by GC regularization.

**ISO for T1–T2.** Encoder-only ISO is the safer choice for multimodal registration. For T1–T2, both ISO-full and ISO-fe improve Dice over SITReg-MIND-Aug, but can increase HD95. Compared with ISO-full, ISO-fe provides a better trade-off, achieving larger Dice gains while degrading HD95 less. This supports adapting only the encoder to accommodate unseen intensity profiles while keeping the deformation decoder frozen.

### 3.2 Validation results of the final submission

Our final validation results are shown in Table 5. For reference, we also report SynthSR-based baselines, which synthesize T1-like images from T2-weighted inputs [18] and then apply a monomodal registration model (SITReg or VFA). Our proposed method performs slightly worse than the SynthSR-based approaches on T1–T2 registration, but the gap is small.

Table 5. Registration results on the LUMIR25 validation set.

(a) Aggregated results.

| Methods | Dice ↑ | HD95 ↓ | TRE ↓ | NDV (%) ↓ |
|---|---|---|---|---|
| Final submission | 0.7570 ± 0.0322 | 3.0388 ± 0.2989 | 2.3099 ± 0.2704 | 0.0030 ± 0.0012 |
| SITReg + synthSR | 0.7610 ± 0.0281 | 2.9968 ± 0.3322 | 2.3102 ± 0.2707 | 0.0024 ± 0.0004 |
| VFA + synthSR | 0.7536 ± 0.0260 | 3.1075 ± 0.3320 | 2.4956 ± 0.3783 | 0.0074 ± 0.0050 |

(b) Subset results.

| Methods | Dice(ID) | Dice(OD) | Dice(MM) | HD95(ID) | HD95(OD) | HD95(MM) |
|---|---|---|---|---|---|---|
| Final submission | 0.7816 | 0.7652 | 0.7240 | 3.1095 | 3.1740 | 2.8329 |
| SITReg + synthSR | 0.7816 | 0.7652 | 0.7363 | 3.1096 | 3.1743 | 2.7065 |
| VFA + synthSR | 0.7744 | 0.7545 | 0.7320 | 3.2033 | 3.2963 | 2.8228 |

## 4 Discussion

The proposed method—SITReg backbone with GC/NDV regularization, MIND-based similarity, intensity randomization, and encoder-only ISO—enables multimodal registration despite being trained only on T1-weighted images. Overall, the recipe is simple, effective and robust. SITReg-MIND-Aug and both ISO variants consistently outperform SynthMorph, a strong contrast-invariant baseline that generalizes by training on synthetically generated label maps and images spanning diverse contrasts and shapes [17]. These gains could be attributed to the proposed real-image–anchored intensity randomization and/or to the strong registration-specific inductive biases (e.g., multi-resolution pyramids, inverse/group consistency) embedded in the SITReg framework. A more controlled ablation is needed to isolate the contribution of each factor. On the T1–T2 validation subset, the proposed method is slightly inferior to synthSR-based baselines. Still, it may be more robust when synthesis fails or hallucinates contrast, since our pipeline does not rely on an explicit synthesis model or a fixed intensity mapping. Overall, our study provides a practical step toward a "registration foundation model" for single-site (brain MRI) multi-modal registration, where a single training source can be leveraged to generalize to unseen domain shifts.

However, there remains a clear accuracy gap across in-domain, out-of-domain, and multimodal validation, indicating room for improvement. A better-designed augmentation scheme that includes local varying effects (e.g., bias fields, local contrast changes, and noise characteristics) and more realistic contrast mimicking may narrow this gap without introducing additional complexity of synthesis. In addition, correlation-based matching remains a promising direction for improving correspondence in challenging multimodal cases, but realizing its benefits at useful scales will require addressing memory constraints.

The final test results confirm that our submitted solution performs strongly for all T1–T1 registration tasks (both in-domain and out-of-domain), but falls short on multi-modal tasks. The strong performance on T1–T1 tasks is attributed to the successful strategies in SITReg. Using MIND loss as we experimented or MIND features as in [19] can improve robustness to out-of-domain contrast variations, but may slightly reduce contour-matching metrics in this particular challenge. Nevertheless, these strategies may be beneficial for more diverse clinical scenarios. We also observe that MIND improves landmark matching (TRE) due to its structure-driven nature.

We further found that the effectiveness of ISO for bridging domain gaps, as reported in [16], needs further verification. In particular, ISO did not improve performance on the out-of-domain high-field validation subset, and we achieved strong out-of-domain T1–T1 results on diverse test sets without ISO. We suspect that pairwise ISO can overfit to intensity matching and partially undermine the benefits of regularization. Understanding when ISO becomes beneficial (i.e., how large a domain gap is needed and under what regularization scheme) remains an open question.

Finally, we want to highlight **the importance of registration-specific inductive biases**, including multi-resolution pyramids, inverse consistency, group consistency, topological preservation or diffeomorphism, and correlation-assisted correspondence establishment. *These ideas are not new, yet they are often overlooked or underexplored in many (if not most) trend-driven learning-based methods.* We also found the correlation-only models very promising. The idea was proposed fairly early in deep learning–based optical flow research [20]. Correlation-only variants can achieve strong results with fewer parameters than intensity-feature-only or hybrid models. Preliminary results (not included) further suggest that correlation-only models may require less training data and be less prone to overfitting to the similarity objective: in our experiments, intensity-feature–based models degrade considerably without regularization, whereas correlation-only models remain comparatively robust.

**Acknowledgments.** This study was funded by NIH (R01CA188300) and DOD (W81XWH2210044).

**Disclosure of Interests.** The authors have no competing interests to declare that are relevant to the content of this article.